\documentclass[prb,twocolumn,amsmath,amssymb,english]{revtex4-1}

\usepackage[dvipdfm]{graphicx}
\usepackage{natbib}
\usepackage{subfigure}
\usepackage{tabularx}
\usepackage{epsfig}
\usepackage{longtable}
\usepackage{amsfonts}
\usepackage{rotating}
\usepackage{subfigure}
\usepackage{amsmath}
\usepackage{hyperref}
\usepackage{babel}
\usepackage{currfile}

\usepackage{color}

\newcommand{\bea}{\begin{eqnarray}}
\newcommand{\eea}{\end{eqnarray}}

\newcommand{\be}{\begin{equation}}
\newcommand{\ee}{\end{equation}}



\begin{document}

\title{Topological crystalline Kondo insulators and universal topological surface states of SmB$_6$}

% \\ \small{(Preliminary draft. Not for distribution)} , file: $\currfilename$)}

\author{Mengxing Ye}
\author{J. W. Allen}
\author{Kai Sun}
\affiliation{Department of Physics, University of Michigan, Ann Arbor, MI 48109, USA}

\date{\today}

%%%%%%%%%%%%%
\begin{abstract}
We prove theoretically that certain strongly correlated Kondo insulators are topological crystalline insulators with 
nontrivial topology protected by crystal symmetries. In particular, we find that SmB$_6$ is such a material. 
In addition to a nontrivial Z$_2$ topological index protected by time reversal symmetry, SmB$_6$ also has nontrival mirror Chern numbers 
protected by mirror symmetries. On the $(100)$ surface of SmB$_6$, the nontrivial mirror Chern numbers 
do not generate additional surface states beyond those predicted by the Z$_2$ topological index. However, on the $(110)$ surface, 
two more surface Dirac points are predicted. Remarkably, we find that for SmB$_6$ both the Z$_2$ topological index and the mirror Chern 
numbers are independent of microscopic details, which enables us to obtain surface state properties that are universal.
\end{abstract}
%%%%%%%%%%%%%

\maketitle

%\tableofcontents

\textit{Introduction---}
In many strongly correlated heavy fermion or mixed valent insulators, hybridization between bands with opposite parities plays a 
very important role in the formation of the insulting 
gap~\cite{Martin1979, Allen1979, Martin1981,Aeppli1992,TSunetsugu1997,Riseborough2000,Coleman2007}. 
It has been shown that this mechanism can result in strongly correlated time-reversal-invariant topological insulators 
(i.e. topological Kondo insulators)~\cite{Dzero2010}. Several candidate materials have 
been predicted theoretically including SmB$_6$  and CeNiSn~\cite{Dzero2010}, CeOs$_4$As$_{12}$ and CeOs$_4$Sb$_{12}$~\cite{Yan2012}, 
and SmS under pressure~\cite{Wolgast2012}. 

Strong supporting evidence has been obtained in recent experimental studies of SmB$_6$. Transport data reveals that 
this material has an insulating bulk with a very robust metallic surface ~\cite{Wolgast2012,Botimer2012}. Quantum oscillations find Fermi surfaces and Dirac points on the $(100)$ and $(110)$ surfaces~\cite{Li2013}, indicating that the metallic surface states are not due to 
trivial mechanisms such as band bending. Fermi pockets on the $(100)$ surface have also been observed in angle-resolved photoemission 
spectroscopy (ARPES)~\cite{Xu2013,Neupane2013,Jiang2013} and the locations of these pockets follow exactly the prediction of the topological 
theory and band structure calculations~\cite{Lu2013}.  Disorder effects have also been investigated. While transport properties show little response 
to nonmagnetic impurities, magnetic impurities greatly suppress the surface conductivity~\cite{Kim2013}, consistent with the topological theory.
Weak-antilocalization has also been reported in the study of magnetorestance~\cite{Thomas2013}.

Despite their strong coupling nature, the low-energy physics of mixed valent materials (e.g. SmB$_6$) can often be described by a 
band structure theory.  However, it is worthwhile to emphasize that this is a low-energy 
effective theory with fermions that emerge from many-body correlations in the f-shell~\cite{Martin1979, Martin1981}.
For SmB$_6$
the bulk band structure has not yet been fully understood. For example it is known that near the Fermi energy ($E_F$) there are three nearly degenerate $f$-bands.  However they have not been 
resolved experimentally and it is still unclear which of them is responsible for opening the insulating gap and the nontrivial topology~\cite{Dzero2010,Dzero2012,Takimoto2011,Lu2013,Xu2013,Neupane2013,Jiang2013}. 
%Although various possibility have been explored in recent theoretical studies~\cite{Dzero2010,Dzero2012,Takimoto2011,Lu2013}, this open question is currently a major challenge.

In this letter, instead of assuming a specific band structure, we focus on universal properties that are independent of 
the yet unknown microscopic details. 
%We find that in addition to time reversal symmetry, point group symmetry also plays an essential role. 
%In particular, we 
We find that on the surface of SmB$_6$, there are two different types of Dirac points, some of which are protected by the time-reversal symmetry, 
while the others are due to lattice symmetries, i.e. SmB$_6$ is not only a topological insulator, but is also a topological crystalline insulator~\cite{Fu2011,Hsieh2012,Xu2012,Dziawa2012,Fang2012a,Tanaka2012,Fu2012,Wang2013,Liu2013,Okada2013} with nontrivial 
mirror Chern numbers. The mirror Chern number is a topological index protected by the point group symmetry of the crystal and it has been 
used in the study of the surface states of various weakly correlated materials  e.g., Bi$_{1−x}$Sb$_x$~\cite{Teo2008}.
It is important to note that in topological insulators with mirror symmetry, it is pretty common to have nontrivial mirror Chern numbers and 
these topological indices don't necessarily result in any addition surface states beyond the prediction of the Z$_2$ topological indices. 
However, for SmB$_6$ the nontrivial mirror Chern number indeed plays a very important role in understanding the surface states and it predicts for the $(110)$ surface two additional surface Dirac points, beyond those predicted by the Z$_2$ topological index.  Because these two surface Dirac points 
are protected by the point group symmetry of the crystal, materials of this type are known as topological crystalline insulators~\cite{Fu2011,Hsieh2012,Xu2012,Dziawa2012,Fang2012a,Tanaka2012,Fu2012,Wang2013,Liu2013,Okada2013}.

It is also worthwhile to emphasize that although we use SmB$_6$ as an example, our techniques and conclusions can easily be generalized to other Kondo insulators, as well as other strongly- or weakly-correlated insulators. In general, for systems with two-fold rotational symmetries, our technique shows that the parity of the mirror Chern number always coincides with the Z$_2$ weak topological index and so does not provide any additional information beyond the Z$_2$ topological indices. However, for systems with higher rotational symmetries (e.g. four-fold), our technique can be used to identify additional surface states beyond those protected by time-reversal symmetry.

\begin{figure}%[htb!]
  \centering
  \subfigure[]{\includegraphics[width=0.4\linewidth]{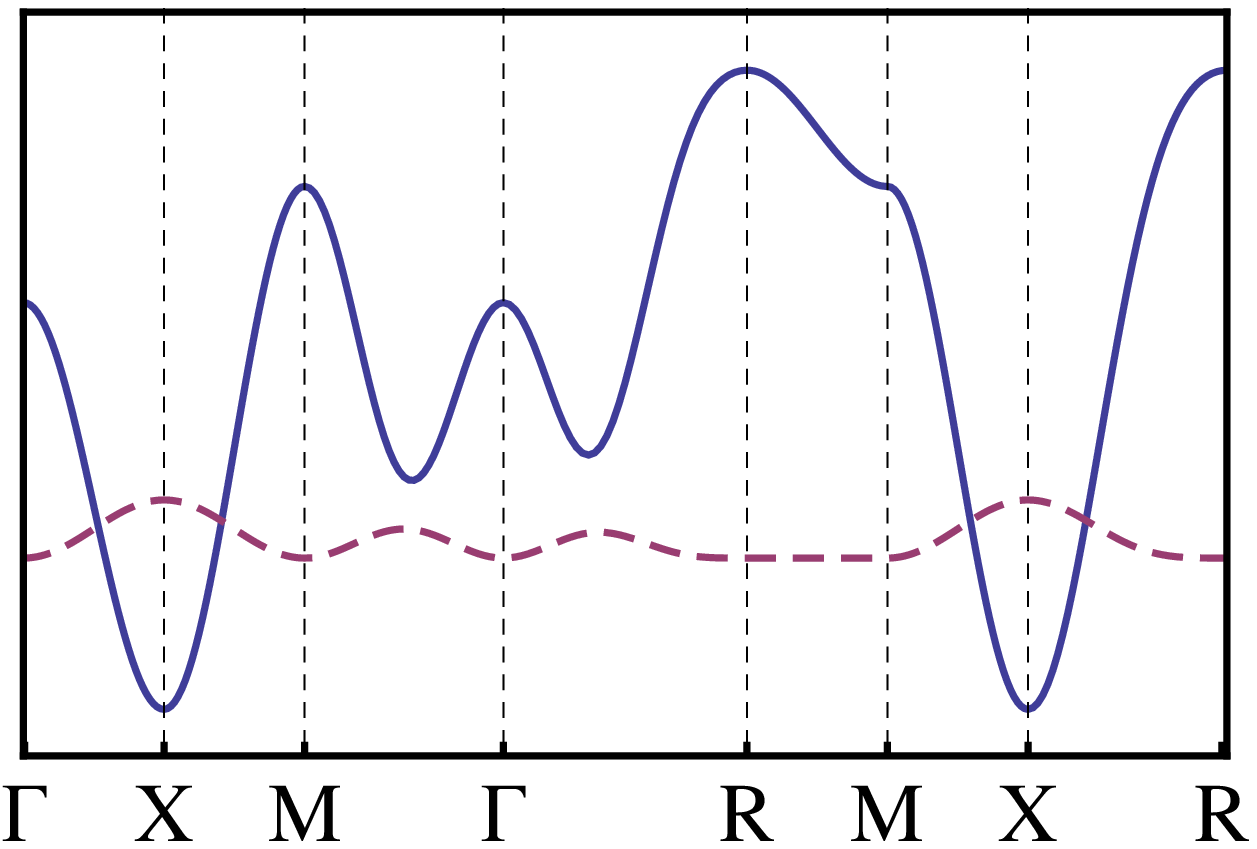}}
  \subfigure[]{\includegraphics[width=0.4\linewidth]{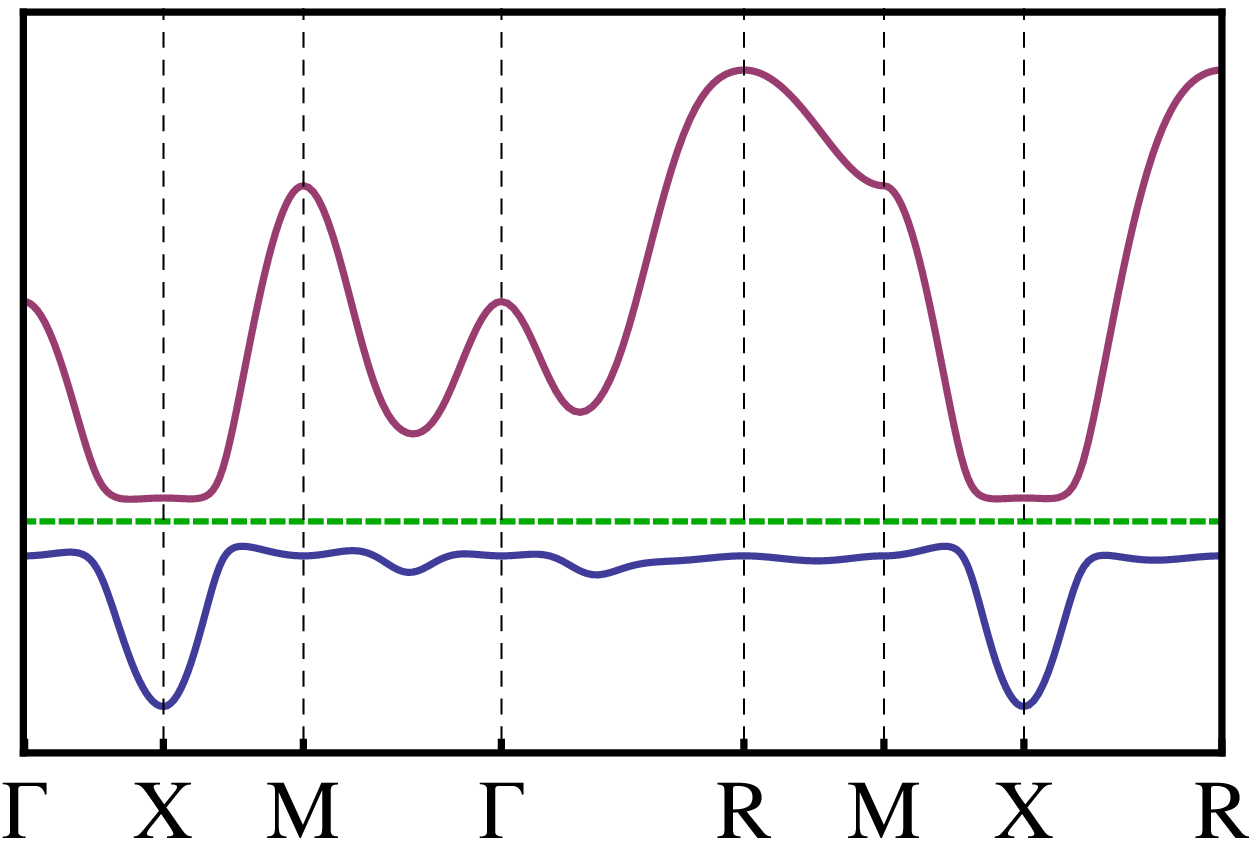}}
%  \subfigure[]{\includegraphics[width=0.3\linewidth]{inversion_3.eps}}
  \caption{Schematic band structures with band inversion at $X$. Here, we only show the two inverted bands, one of which has odd party 
(dashed line in a) and the other has even parity (solid line in a). In Fig. (a), hybridization between the two bands are set to zero.
In Fig. (b), band hybridization is introduced, which opens an insulating gap. The green dotted line marks $E_F$.
At $X$, the lower (upper) band has even (odd) parity, where at all other high symmetry points, the lower (upper) band has odd (even) parity.}
\label{fig:band_inversion}
\end{figure}

\textit{Band structure of SmB$_6$---} %In this section, we will review only the aspects of the SmB$_6$ electronic structure that are essential for our conclusions. Additional details and open questions that may arise in any of our proofs will be addressed in later parts.
In SmB$_6$, the Sm atoms form a simple cubic lattice and inside each cube there is a tetrahedron of six B atoms. From band calculations 
and spectroscopic experiments it has long been known that the relevant bands near $E_F$ come from the $4f$ and $5d$ states of Sm. Over most of the Brillouin zone the $5d$-bands are above $E_F$ while the $4f$-bands stay below. However, around the $X$ points 
[$(\pi,0,0)$, $(0,\pi,0)$ and $(0,0,\pi)$], the energy of one of the $5d$ bands goes below $E_F$ and in order to maintain the number 
of valance bands below $E_F$, which is necessary for an insulator, one of the $4f$ band must go above. Hybridization between these two bands opens an insulating gap as shown in Fig.~\ref{fig:band_inversion}. This phenomenon is known as a \textit{band inversion}, which played a very important role in the discovery of %time-reversal invariant topological insulators
the quantum spin Hall insulators~\cite{Bernevig2006}. Because $d$-states ($f$-states) have even (odd) parity, if we choose Sm as the inversion center, the two inverted bands must have opposite parities at $X$. In this Letter, all our conclusions rely on only two assumptions: (1) two bands with opposite parities are inverted at $X$ and (2) the material has an insulating bulk.

%%%%%%%%%%%%%%%%%%%%%%%%%%%%%%%%%%%%%%%%%%%%%%%%%%%%%%%%%%%%%%%%%%%%%%%%%%%%%%%%%%%%%%%%

\textit{Time-reversal symmetry and the Z$_2$ topological index---}
The Z$_2$ topological index for Kondo insulators (including SmB$_6$) has been computed before by one of the authors and coworkers~\cite{Dzero2010,Dzero2012}.  Here we obtain this topological index using a different approach in order to demonstrate a construction which will be used later to compute the mirror Chern number.
In Ref.~\onlinecite{Fu2007} it was proved that in a material with time reversal and space inversion symmetries, the Z$_2$ topological index can be computed using parity eigenvalues of the valence bands at high symmetry points.
\bea
(-1)^\nu=\prod_{m=1}^{N}\prod_{i=1}^{8}\xi_{m}(\Gamma_i),
\label{eq:STI}
\eea
where $\nu$ is the strong topological index; $m$ is the band index for the valence bands, each of which is required by symmetry to be doubly-degenerate; $\Gamma_i$ represents the eight high symmetry points in the Brillouin zone ($\Gamma$, $X$, $M$ and $R$ for 
a cubic lattice); and $\xi_{m}(\Gamma_i)=\pm1$ is the parity eigenvalue of band $m$ at $\Gamma_i$.

Instead of computing the product directly, here we obtain the topological index $\nu$ by comparing the band structure of SmB$_6$ with that of a trivial insulator, which is obtained by raising (lowering) the energy of the $d$ ($f$) bands such that the band inversion is eliminated. By putting the Fermi energy between the $f$ and $d$ bands, the system is obviously a trivial insulator (analogous to the insulating state of single valent SmS at ambient pressure).
For both SmB$_6$ and this trivial insulator the Z$_2$ topological indices ($\nu_{\textrm{SmB}_{6}}$ and 
$\nu_{\textrm{trivial}}$) can be formulated using Eq. \eqref{eq:STI}. Here we focus on the ratio
\begin{align}
\frac{(-1)^{\nu_{\textrm{SmB}_{6}}}}{(-1)^{\nu_{\textrm{trivial}}}}=\prod_{i=1}^{8}\prod_{m=1}^{N}\frac{\xi^{\textrm{SmB}_6}_{m}(\Gamma_i)}{\xi^{\textrm{trivial}}_{m}(\Gamma_i)}.
\label{eq:STI_ratio_1}
\end{align}
On the l.h.s., the denominator is the identity %because for a trivial insulator $\nu_{\textrm{trivial}}=0$. 
($\nu_{\textrm{trivial}}=0$). For the r.h.s., it is easy to see that because these two insulators have exactly the same (conduction and valence) bands except for those involved in band inversions, only %the latter of the parity eigenvalues fail to cancel.  After these simplifications, we get
the parity eigenvalues of the inverted bands fail to cancel.  After these simplifications, we get
\begin{align}
(-1)^{\nu_{\textrm{SmB}_{6}}}=\left[\frac{\xi^{{\textrm{SmB}_{6}}}_{\textrm{IV}}(X)}{\xi^{{\textrm{trivial}}}_{\textrm{IV}}(X)}\right]^3
=\left[\frac{\xi^{{\textrm{SmB}_{6}}}_{\textrm{IV}}(X)}{\xi^{{\textrm{SmB}_{6}}}_{\textrm{IC}}(X)}\right]^3=-1.
\label{eq:STI_ratio_2}
\end{align}
The power three arises because there are three $X$ points. $\xi^{{\textrm{SmB}_{6}}}_{\textrm{IV}}(X)=+1$ and $\xi^{{\textrm{SmB}_{6}}}_{\textrm{IC}}(X)=-1$ 
are the parity eigenvalues of the inverted-valence (IV) and inverted-conduction (IC) bands in SmB$_6$ at $X$ respectively, where the former
comes from parity even ($+1$) $5d$-states and the latter are due to parity odd ($-1$) $4f$-states. $\xi^{{\textrm{trivial}}}_{\textrm{IV}}(X)$ is the parity eigenvalue
of the corresponding valence band in the trival insualtor. Here we used the fact that at $X$ the inverted conduction band in SmB$_6$ is identical to the corresponding valence band 
in the trivial insulator: 
$\xi^{{\textrm{SmB}_{6}}}_{\textrm{IC}}(X)=\xi^{{\textrm{trivial}}}_{\textrm{IV}}(X)$, which enables us to substitute the denominator into 
$\xi^{{\textrm{SmB}_{6}}}_{\textrm{IC}}(X)$. The final result ($-1$) indicates that SmB$_6$ is a strong topological insulator.

This technique can be easily generalized to other materials with time-reversal and space-inversion symmetries and can also be utilized to 
compute the weak topological indices (See SI for details). 

\textit{Mirror Chern number---}
For insulators with a mirror symmetry we can define another topological index, the mirror Chern number. We start our discussion by considering 2D insulators with D$_{nh}$ symmetry.   The mirror symmetry that protects the nontrivial topological structure is horizontal reflection in a plane parallel to the 2D surface.  
%For simplicity, we focus on the D$_{4h}$ group, but the conclusions can easily be generalized. 
For particles with half-integer spins, we can use the mirror eigenvalues of the horizontal mirror plane to classify all Bloch waves into two groups: $|\psi_m^{+}(\mathbf{k})\rangle$ and $|\psi_m^{-}(\mathbf{k})\rangle$,
where the ``$+$" (``$-$") states have mirror eigenvalue $+i$ ($-i$). The subscript $m$ is the 
band index and $\mathbf{k}$ is the momentum. Using either $|\psi_m^{+}(\mathbf{k})\rangle$ or $|\psi_m^{-}(\mathbf{k})\rangle$, we can define the mirror Chern number $C^+$ or $C^-$ as
\begin{align}
C^{\pm}=i \epsilon_{ab} \sum_{m=1}^{N} 
\int_{BZ} \frac{d\mathbf{k}}{2\pi}\langle \partial_{a}  \psi _m^{\pm}(\mathbf{k})|\partial_{b}  \psi _m^{\pm}(\mathbf{k}) \rangle.
\end{align}
Here, we sum over all valence bands ($m=1$, $\ldots$, $N$) and integrate over the whole Brillouin zone. $\epsilon_{ab}$ is the 2D Levi-Civita symbol with $a$ and $b$ being $k_x$ and $k_y$.   
The sum $C=C^++C^-$ is the (first) Chern number.  %, which describes the quantum Hall effect. 
Because $C$ is odd under the mirror reflection, the point group symmetry requires $C=0$ , i.e. $C^+=-C^-$. 

The mirror Chern number has properties similar to %the more well-known topological index, 
the first Chern number~\cite{Teo2008,Hsieh2012}. 
 If the mirror symmetry is preserved both in the bulk and 
on the edge, a nonzero mirror Chern number implies the existence of (left- or right- moving) edge modes, and the mirror Chern number $C^+$ ($C^-$) is the difference between the numbers of left-moving and right-moving edge modes with mirror eigenvalue $+i$ ($-i$). 
Because modes with opposite mirror eigenvalues cannot hybridize, these edge states cross each other and form Dirac points on the edge.
These Dirac points are protected by the mirror symmetry, without which $|\psi_m^{+}(\mathbf{k})\rangle$ will hybridize with $|\psi_m^{-}(\mathbf{k})\rangle$ and gap out the Dirac points.

In a 3D insulator, the mirror Chern number can be defined for any 2D planes with horizontal mirror symmetry in the 3D Brillouin zone.
%We first study the $k_z=0$ plane of SmB$_6$, which has symmetry $D_{4h}$. 
For example, the $k_z=0$ plane of SmB$_6$ has symmetry D$_{4h}$ (Here, the main axis directions
are chosen to be along the three four-fold-rotational axes of the cubic lattice). In Ref.~\onlinecite{Fang2012}, 
it is proved that for a system with $n$-fold rotational symmetry (C$_n$), the Chern number can be computed (up to modulo $n$) 
as a product of eigenvalues of the rotational operators 
at high symmetry points. 
Here we use the same principle to compute the mirror Chern number $C^+$ 
%by focusing on the quantum states $|\psi_m^{+}(\mathbf{k})\rangle$,
\begin{align}
(i)^{C^{+}}=\prod_{m=1}^{N}(-1) \eta_{m}(\Gamma)\eta_m(M)\zeta_{m}(X),
\label{eq:MCI_C4}
\end{align}
where $\eta_{m}(\Gamma)$ and $\eta_m(M)$ are eigenvalues of the $90^\circ$-rotation along the normal direction of the 2D plane ($z$) 
for band $|\psi^+_m(\mathbf{k})\rangle$ at $\Gamma$ and $M$ respectively. 
$\zeta_m(X)$ is the eigenvalue of the $180^\circ$-rotation along the same axis at $X$. The band index $m$ runs over all valence bands. 
It is important to notice that we require 
$|\psi^+_m(\mathbf{k})\rangle$ to be eigenstates of the mirror operator (about the $x-y$ plane), 
as well as the four-fold and two-fold rotational operators (along $z$). This can always be achieved 
because these operators commute with one another.

We compute this mirror Chern number by comparing SmB$_6$ with the trivial insulator discussed above
\begin{align}
\frac{(i)^{C^{+}_{\textrm{SmB}_6}}}{(i)^{C^{+}_{\textrm{trivial}}}}
=\prod_{m=1}^{N}\frac{\eta_{m}^{\textrm{SmB}_6}(\Gamma)\eta_m^{\textrm{SmB}_6}(M)\zeta_{m}^{\textrm{SmB}_6}(X)}
{\eta_{m}^{\textrm{trivial}}(\Gamma)\eta_m^{\textrm{trivial}}(M)\zeta_{m}^{\textrm{trivial}}(X)}.
\label{eq:mirror_C_ratio}
\end{align}
Similar to the Z$_2$ topological index, most of the eigenvalues on the r.h.s. cancel out, except for those of the inverted bands at $X$, %and therefore
\begin{align}
(i)^{C^{+}_{\textrm{SmB}_6}}=\frac{\zeta_{\textrm{IV}}^{\textrm{SmB}_6}(X)}{\zeta_{\textrm{IV}}^{\textrm{trivial}}(X)}
=\frac{\zeta_{\textrm{IV}}^{\textrm{SmB}_6}(X)}{\zeta_{\textrm{IC}}^{\textrm{SmB}_6}(X)}=-1.
\label{eq:mirror_C_ratio_2}
\end{align}
Here we use the fact that $C^{+}_{\textrm{trivial}}=0$ and IC and IV represent inverted-conduction and inverted-valence bands respectively.
Same as in Eq.~\eqref{eq:STI_ratio_2}, we substitute the denominator using the fact 
$\zeta_{\textrm{IV}}^{\textrm{trivial}}(X)=\zeta_{\textrm{IC}}^{\textrm{SmB}_6}(X)$

The ratio in Eq.~\eqref{eq:mirror_C_ratio_2} can be easily determined using the representations of the space group. At a momentum point, all the bands 
can be labeled according 
to their symmetry properties using double group representations of the little group at this momentum point~\cite{Dresselhaus2010}. 
For $X$, the little group is D$_{4h}$, 
and the group theory requires that all the parity even (odd) $d$- ($f$-) bands must have either the symmetry of $\Gamma_6^{+}$ 
or $\Gamma_7^{+}$ ($\Gamma_6^{-}$ or $\Gamma_7^{-}$). Here, the $+/-$ sign represents even/odd parity. 
As proved in the Supplementary Information (SI), for the two inverted bands, 
because the hybridization between them opens up a full gap, they must belong to the same representation up to their differing
parities~\cite{Koster1963,Dresselhaus2010}. 
%This is because the compatibility relation dictates that if one of the band has symmetry $\Gamma_6$ at $X$ while the other band has symmetry $\Gamma_7$, the hybridization between them must vanish along the main axis of the k-space ($k_x$, $k_y$ or $k_z$, which are typically labeled as $\Delta$ in space group literature), and thus the insulating gap must clap along $\Delta$, which is in contradictory to transport and ARPES data. 
%If both the inverted bands have symmetry $\Gamma_7$ (or $\Gamma_6$) but with opposite parities, 
This conclusion is enough to determine the ratio in Eq.~\eqref{eq:mirror_C_ratio_2} using eigenvalues provided by the group 
representations, which is $-1$ as shown in SI. This result tells us that the mirror Chern number $C^{+}_{\textrm{SmB}_6}=2$ up to modulo $4$, 
and therefore  SmB$_6$ is a topological crystalline insulator.

Using the same technique, we find that the mirror Chern number of the $k_z=\pi$ plane is $1$. For mirror planes with D$_{2h}$ 
symmetry (e.g. the $k_x=k_y$ plane), our technique can be used to show that the parity of the mirror Chern number always coincides 
with the weak topological index, implying no new insight beyond the Z$_2$ indices (See SI for details).

\begin{figure}
  \centering
  \subfigure[]{\includegraphics[width=0.3\linewidth]{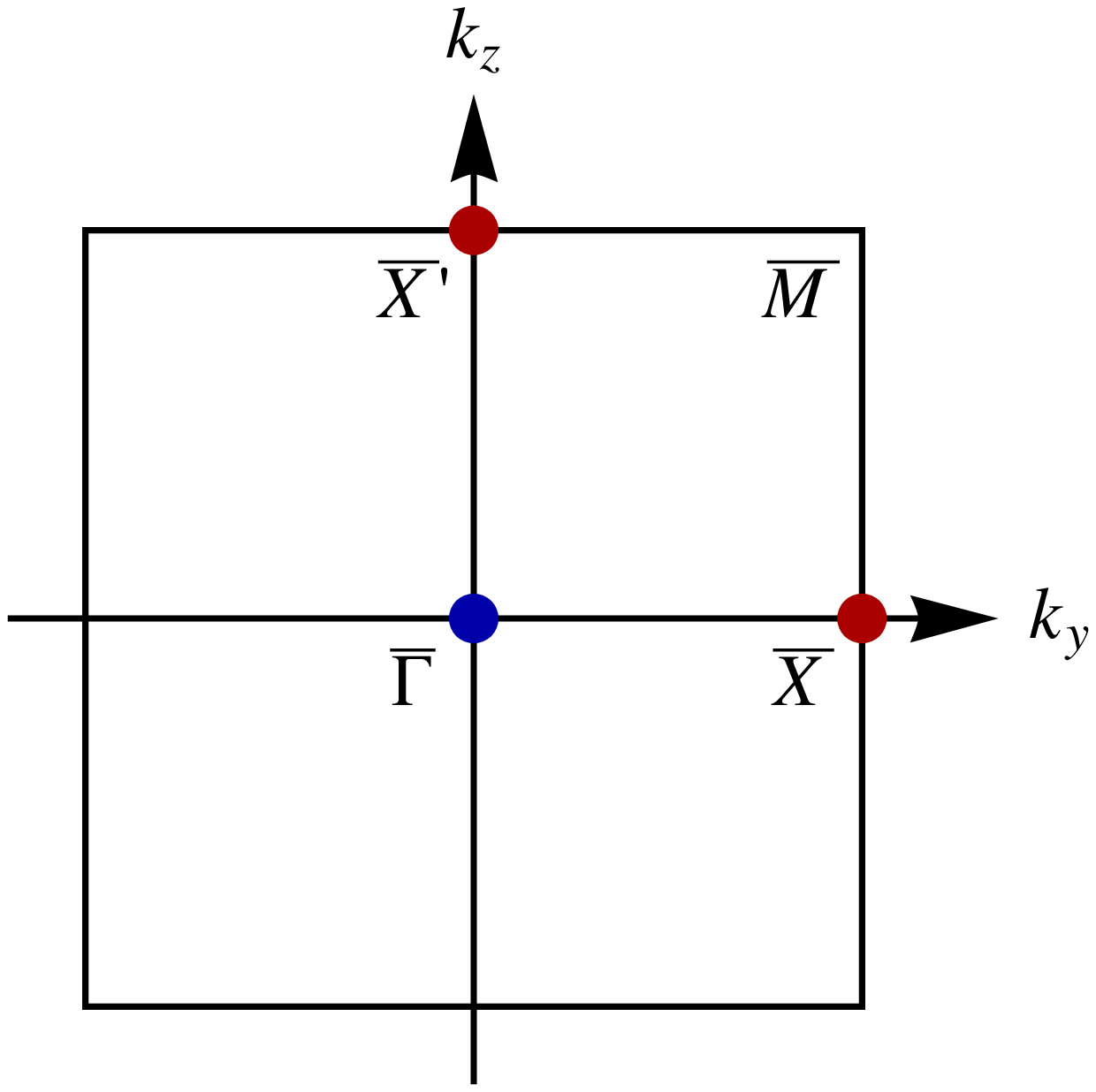}}
  \subfigure[]{\includegraphics[width=0.3\linewidth]{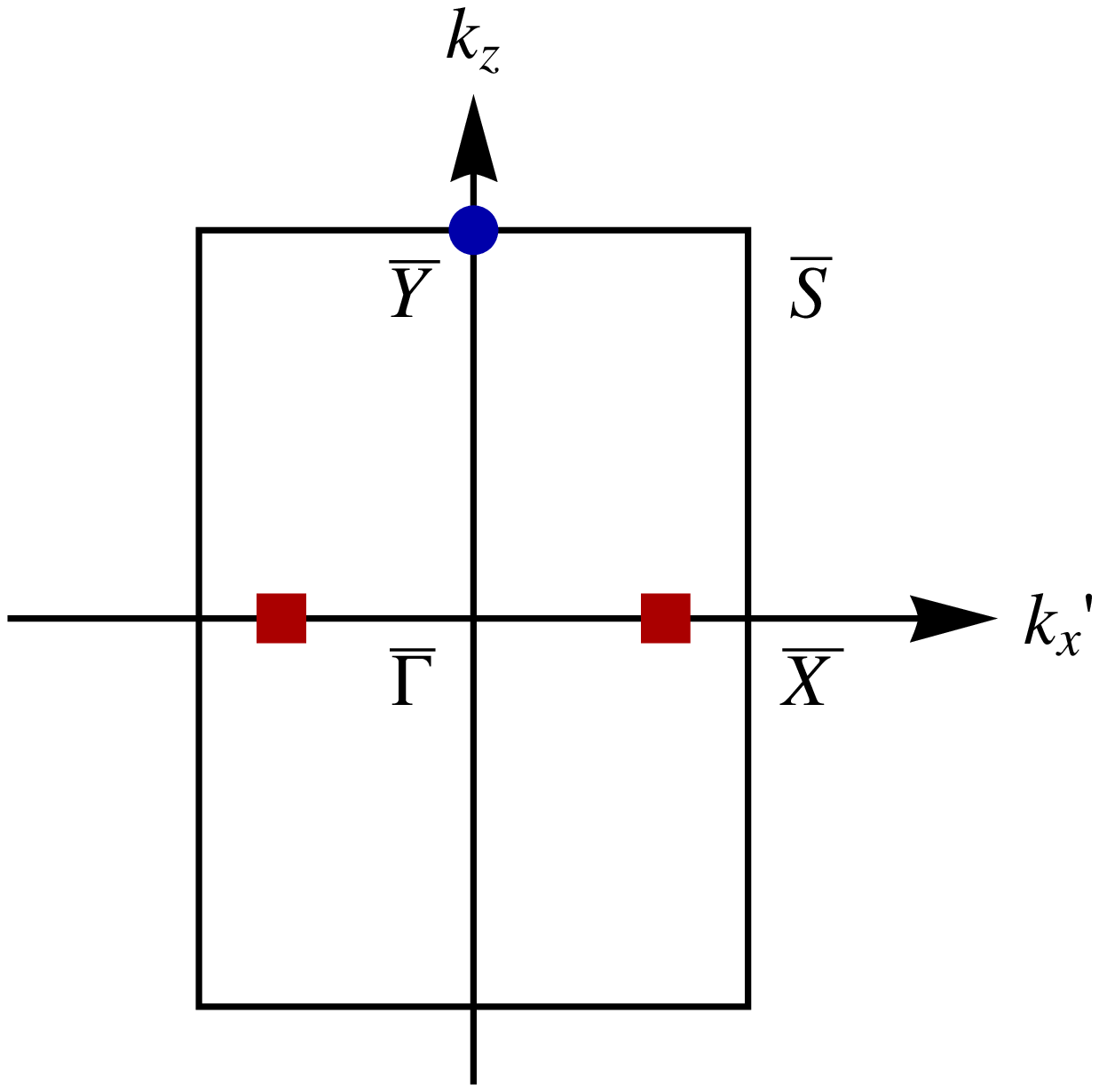}}
  \subfigure[]{\includegraphics[width=0.3\linewidth]{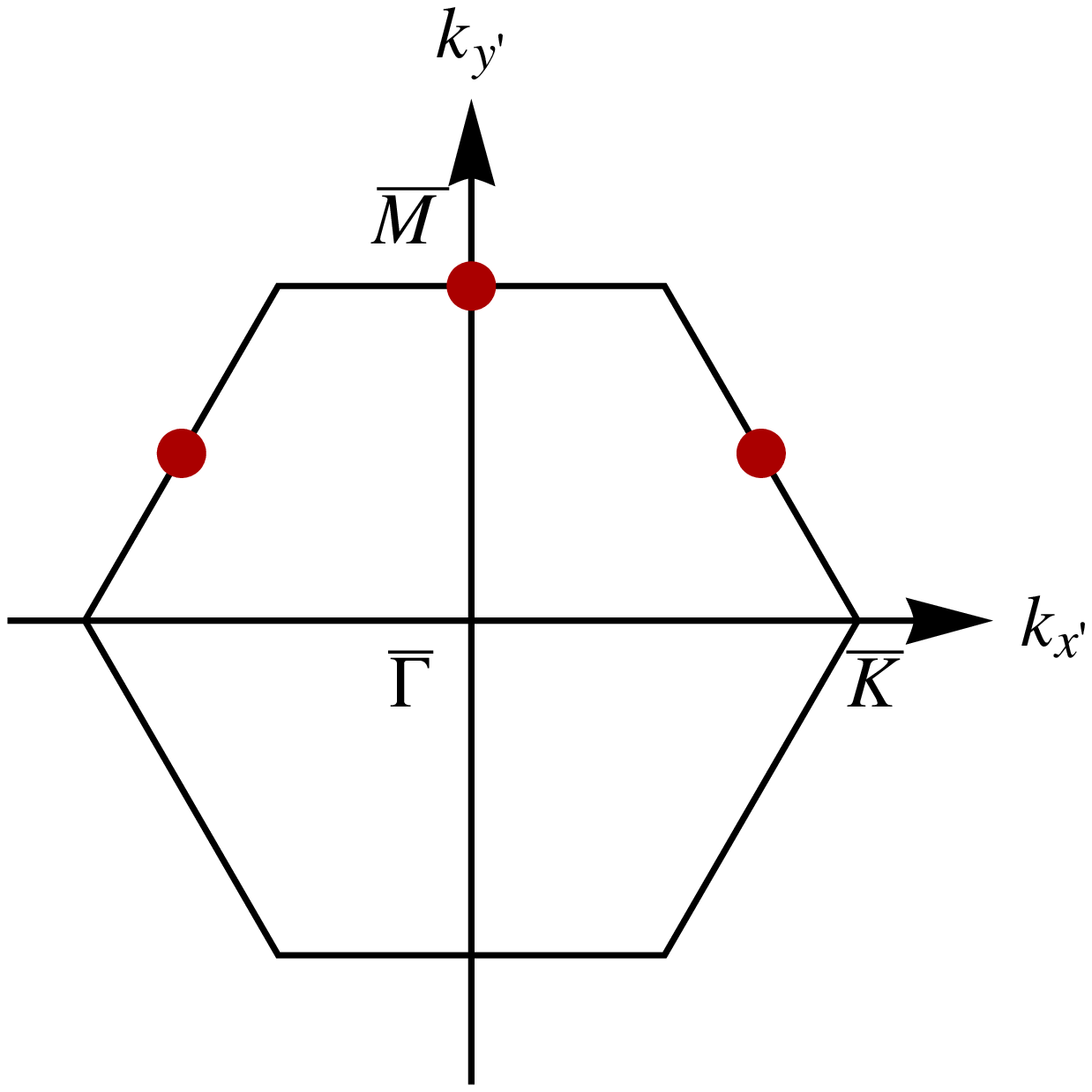}}
  \caption{Surface Dirac points on the (a) $(100)$, (b) $(110)$, and (c) $(111)$ surfaces of SmB$_6$. Disks (squares) represent Dirac points protected 
by the time-reversal (mirror) symmetry.} 
%On the $(100)$ surface, 
%all the three Dirac points (disks) are protected by the time-reversal symmetry. On the $(110)$ surface, the Dirac point at $\bar{Y}$ (disk)
%is protected by the time-reversal symmetry, while the other two (squares) are due to the nontrivial mirror Chern number and are protected
%by the crystal symmetry.}
\label{fig:surface}
\end{figure}

\textit{Surface states---} The topological indices discussed above have direct impacts on the surface states. For the  Z$_2$  
topological index, if nonzero, it predicts Dirac points at certain high symmetry points of the surface Brillouin zone. 
For a nontrivial mirror Chern number, it results in Dirac points along certain high symmetry line in the surface Brillouin zone. 

For the Z$_2$ index, the surface Dirac cone can be found by projecting the bulk high symmetry points with band inversions ($X$ for SmB$_6$) onto the 
surface~\cite{Fu2007}. If an odd number of the bulk $X$ points are projected to a surface high symmetry point, there must be a surface Dirac point at this location. 
For the $(100)$ surface of SmB$_6$, this technique predicts three Dirac points at $\bar{\Gamma}$, $\bar{X}$ and $\bar{X}'$. 
On the $(110)$ surface, one of the bulk X points is projected to $\bar{Y}$, while the other two are both projected to $\bar{X}$. As a result,
 time-reversal symmetry only protects one Dirac point on the $(110)$ surface located at $\bar{Y}$ (Fig.~\ref{fig:surface}).

For the mirror Chern number defined on a 2D plane in the 3D Brillouin zone, if this 2D plane is perpendicular to the surface, its projection on 
the surface Brillouin zone forms a 1D line, along which there should be (at least) $|C^+|$ surface Dirac points.
On the $(100)$ surface of SmB$_6$, mirror Chern numbers give us no additional Dirac points. 
For example, the bulk $k_z=0$ plane (with mirror Chern number $2$) is projected to the $k_z=0$ line onto the (100) surface. The mirror Chern number 2 implies that there should be two left-moving modes with parity $+i$ and 
two right-moving modes with parity $-i$ along this surface high symmetry line, and the intersections between these modes will form 
two surface Dirac points. By comparing with Fig.~\ref{fig:surface}(a), it is easy to notice that these two surface Dirac points are 
just the Dirac points at $\bar{\Gamma}$ and $\bar{X}$ which are required
by time-reversal symmetry. The bulk $k_z=\pi$ plane with mirror Chern number $1$ is projected to the $k_z=\pi$ line on the $(100)$ surface. 
The mirror Chern number requires this surface line to have one Dirac point, which is just the Dirac point at $\bar{X}'$. 
The bulk $k_z=k_y$ plane also has  mirror Chern number $1$, and therefore the surface $k_z=k_y$ line should have one Dirac point, which is 
the $\bar{\Gamma}$ Dirac point. 

For the $(110)$ plane [Fig.~\ref{fig:surface}(b)], however, the mirror Chern number of the bulk $k_z=0$ plane predicts two additional surface Dirac points. By projecting
this plane onto the $(110)$ surface, we obtain the $k_z=0$ line and the mirror Chern number requires that there should be two Dirac points along
this line. In the same time, the Z$_2$ topological index generates no surface Dirac points along this line, which implies that these two Dirac
points are beyond the prediction of the Z$_2$ topological index and are not protected by time-reversal symmetry (but by the mirror symmetry).
The two-fold rotational symmetry requires these two Dirac points to be located at $(k,0)$ and $(-k,0)$ on the surface $k_x'$-$k_z$ plane, 
as marked in Fig.~\ref{fig:surface}(b). These two Dirac points are the direct prediction of the topological crystalline Kondo insulator and can be 
verified directly in experiments.

For other surfaces, e.g. (111), if the mirror planes perpendicular to the surface only have D$_{2h}$ (or lower) symmetry, 
this technique of computing mirror Chern numbers doesn't offer any new insight beyond the Z$_2$ topological 
indices. It is also worthwhile to emphasize that in addition to the universal part of the surface states predicted above, other mechanism could 
induce extra pockets on the surface, e.g. band bending, which relies on microscopic details and is beyond the scope of this investigation.  

\begin{figure}
  \centering
\includegraphics[width=0.8\linewidth]{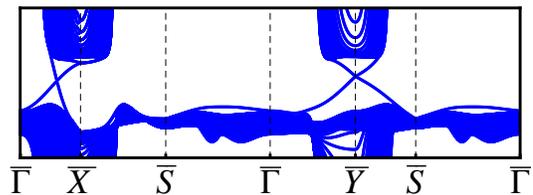}
 \caption{Surface states on the $(110)$ surface.
The plot shows the band structure of our minimum model with open a $(110)$ surface. Surface Dirac points emerge inside the band gap. 
The Dirac point at $\bar{Y}$ is protected by the time-reversal symmetry, where the other (located between $\bar{\Gamma}$ and $\bar{X}$) 
is one of the two Dirac points protected by the mirror symmetry.}
\label{fig:model}
\end{figure}

\textit{Model calculation---} To properly model the band structure of SmB$_6$ is highly nontrivial. This material has three nearly 
degenerate $f$-bands around $E_F$, which form a $j=5/2$ representation of the $SU(2)$ group. The space group theory tells us that at 
the $X$ point, two of these three band shall have $\Gamma_7^-$ symmetry, while the other has $\Gamma_6^-$. Experimentally, because they are 
very close in energy, these three bands have not been clearly resolved and thus we don't know exactly which band gets inverted.
Knowledge of the $d$-band is relatively clear because there is only one $d$-band near $E_F$. Early band calculations without spin-orbit coupling show that in similar materials the relevant 
$5d$-band has $\Gamma_3^+$ symmetry at $X$~\cite{Hasegawa1977}. 
If we include spin-orbit coupling, the symmetry for this band is $\Gamma_3^+\otimes\Gamma_6^+=\Gamma_7^+$ ~\cite{Martin1979, Martin1981} consistent with later numerical studies~\cite{Lu2013}. 
Because SmB$_6$ has a full gap, as described above, the space group symmetry requires the inverted $f$-band to 
have the symmetry of $\Gamma_7^-$ at $X$, in order to match the symmetry of the $d$-band. However, which one of the two $\Gamma_7^-$ bands is inverted is unclear. One addition complication comes 
from the cubic symmetry, which distinguishes these two $\Gamma_7^-$ $f$-bands. If we follow the three $f$-bands from $X$ to $\Gamma$, the group 
theory requires one of the $\Gamma_7$ band to merge with the $\Gamma_6$ band at $\Gamma$ forming a quartet ($\Gamma_8^-$), where the other $\Gamma_7$ band remains a doublet ($\Gamma_7^-$). Whether the inverted $\Gamma_7$ band is part of the quartet at $\Gamma$ or it comes from the doublet at 
$\Gamma$ is still unclear~\cite{Dzero2010,Dzero2012,Takimoto2011,Lu2013,Xu2013,Neupane2013,Jiang2013}.

Because we have proved that all the topological properties are independent of these microscopic details, 
the topological surface states of SmB$_6$ can be obtained (qualitatively) using a minimum two-band model. The minimum model we 
constructed here is a tight-binding model on a cubic lattice with two orbitals per site with opposite parities. 
By choosing parameters such that the two bands formed by these two orbitals are inverted at $X$, all the necessary ingredients 
are captured and the surface states of this model as shown in Fig.~\ref{fig:model} are in perfect agreement with the universal conclusions proved above.
In addition, we also computed surfaces states on the $(100)$ and $(111)$ surfaces (not shown), which also agree with our universal theory.
 
\textit{Discussion---}
According to the classification of topological band-insulators in Ref.~\onlinecite{Slager2013}, SmB$_6$ belongs to the $T-p3(4)_X$ (XYZ) class. 
For other insulators in this class, they shall share the same qualitative surface states and our technique can be generalize
to obtain the topological surfaces states for insulators in other symmetry classes discussed therein. 
In recent quantum oscillation measurement, a Dirac point on the (110) surface of SmB$_6$ is observed, but it is still unclear which 
of the three Dirac points in Fig.~\ref{fig:surface}(b) is responsible for the observed quantum oscillations. To fully understand the 
$(110)$ surface, experimental techniques with momentum resolution are necessary (e.g. ARPES). 

\textit{Acknowledgment---} K.S. thanks Liang Fu for constructive comments. The work was supported in part by the MCubed program at
the University of Michigan.

\appendix

\section{Symmetry of the inverted bands}
For SmB$_6$, at the $X$ point where the band inversion takes place, the two inverted bands can be labeled according to the group representations
of the little group at $X$ (D$_{4h}$). For spin-$1/2$ particles, the representation must be one of the four double group representations 
$\Gamma_6^+$, $\Gamma_6^-$, $\Gamma_7^+$, or $\Gamma_7^-$.

%This is because the compatibility relation dictates that if one of the band has symmetry $\Gamma_6$ at $X$ while the other band has symmetry $\Gamma_7$, the hybridization between them must vanish along the main axis of the k-space ($k_x$, $k_y$ or $k_z$, which are typically labeled as $\Delta$ in space group literature), and thus the insulating gap must clap along $\Delta$, which is in contradictory to transport and ARPES data. 
%If both the inverted bands have symmetry $\Gamma_7$ (or $\Gamma_6$) but with opposite parities, 

As shown in Fig.1, it is the hybridization between the two inverted bands that opens up the insulating gap. This hybridization has strong 
dependence on the symmetry of the inverted bands. Most importantly, the symmetry of the two bands along $\Delta$ 
(i.e. the line connecting $\Gamma$ and $X$ along the main axis direction) dictates whether the hybridization can open a full gap or not, 
i.e. whether or not the band crossing points located between $\Gamma$ and $X$ shown in Fig 1.(a) can be gapped out
(Band crossings away from $\Delta$ don't provide addition information on the symmetry of the bands and thus will not be discussed).

The compatibility relation of the space group~\cite{Koster1963} tells us that if a band has symmetry $\Gamma_6^+$ or $\Gamma_6^-$ at $X$, 
along the $\Delta$ line, this band shall corresponds to the $\Gamma_6$ representation of the little group at $\Delta$ (C$_{4v}$).  Similarly, for a band with symmetry $\Gamma_7^+$ or $\Gamma_7^-$ at $X$, the symmetry of this band along $\Delta$ is $\Gamma_7$.
The parity eigenvalues play no role along $\Delta$, because space inversion is not part of the C$_{4v}$ group.

If one of the two inverted bands belongs to the $\Gamma_6^{\pm}$ representation at $X$, while the other one
is a $\Gamma_7^{\pm}$ band, these two bands will have different symmetries along $\Delta$ (i.e. a $\Gamma_6$ band and a $\Gamma_7$ band
along $\Delta$). This symmetry difference prohibits hybridization between these two bands along $\Delta$. As a result, band crossing point
shown in Fig.1(a) between $\Gamma$ and $X$ cannot be gapped out,
%the bulk hybridization gap must contain nodal points along $\Delta$ 
i.e. the bulk contains symmetry-protected 3D Dirac points somewhere along $\Delta$, and therefore we cannot
 have a fully insulating bulk.

To open an insulating gap in the bulk, the two inverted bands must have the same symmetry at $X$ (up to different parities).  For example, if one of
the inverted bands have symmetry $\Gamma_7^+$ at $X$, while the other has $\Gamma_7^-$, these two bands along $\Delta$ has the same symmetry
$\Gamma_7$ and thus a hybridization gap becomes allowed. Same is true if the two bands have are $\Gamma_6^+$ and $\Gamma_6^-$ bands at $X$.

For SmB$_6$, band structure calculations suggest that the two inverted bands have symmetries $\Gamma_7^+$ and $\Gamma_7^-$ at $X$, but our conclusions
remain valid even if the two bands are $\Gamma_6^{\pm}$.

\section{Z$_2$ topological indices}
Here, we generalize the formula that we used to compute the Z$_2$ topological index of SmB$_6$.
For an arbitrary insulator with time-reversal and space-inversion symmetries, the strong topological index $\nu$ can be computed 
using the following formula, where $(-1)^\nu$ equals the ratio between total parity of the inverted valence bands and that of the inverted 
conduction bands at all high symmetry points.
\begin{align}
(-1)^{\nu}=\prod_{i=1}^{8}\prod_{m_i}\frac{\xi_{m_i}^{\textrm{IV}}(\Gamma_i)}{\xi_{m_i}^{\textrm{IC}}(\Gamma_i)},
\end{align}
where $\Gamma_i$ represents the eight high symmetry points in the 3D Brillouin zone ($i=1, \ldots, 8$) and the product $\prod_{m_i}$ runs over all the inverted bands at the high symmetry point $\Gamma_i$. The superscript IV (IC)
 represents the inverted-valence (inverted-conduction) bands and $\xi$ is the parity eigenvalue for the corresponding band at a
 high symmetry point. %If we apply this formula to SmB$_6$, the formula shown in the manuscript is recovered.

The same technique can be used to compute the weak topological index, if we only use the eigenvalues at four high symmetry
points on a high symmetry plane.
\begin{align}
(-1)^{\nu_{\textrm{weak}}}=\prod_{i=1}^{4}\prod_{m_i}\frac{\xi_{m_i}^{\textrm{IV}}(\Gamma_i)}{\xi_{m_i}^{\textrm{IC}}(\Gamma_i)}.
\label{app:eq:weak_z2}
\end{align}

\section{2D planes with D$_{2h}$ symmetry}
\label{app:sec:D2h}
For a 2D system with two-fold rotational symmetry $C_2$, the parity of the Chern number can be determined using 
eigenvalues of rotational operators at high symmetry points as shown in Ref.~\onlinecite{Fang2012}.
For a 2D system with D$_{2h}$ symmetry, the mirror Chern number (up to modulo $2$) can be determined using a similar technique.
\begin{align}
(-1)^{C^{+}}=\prod_{i=1}^{4}\prod_{m=1}^{N}(-1) \zeta_{m}(\Gamma_i),
\end{align}
where $\Gamma_i$ are the four high-symmetry points of the 2D Brillouin zone. $\zeta_{m}(\Gamma_i)$ is the eigenvalue of the $180^\circ$-rotation 
along the direction normal to the plane for the Bloch state $|\psi^+_m(\mathbf{k}=\Gamma_i)\rangle$. 

If we compute the mirror Chern number by comparing with a trivial insulator with no band inversion, we find that
\begin{align}
(-1)^{C^{+}}=\prod_{i=1}^{4}\prod_{m_i} \frac{\zeta^{\textrm{IV}}_{m_i}(\Gamma_i)}{\zeta^{\textrm{IC}}_{m_i}(\Gamma_i)},
\end{align}
where $\Gamma_i$ represents four high symmetry points of the 2D Brillouin zone ($i=1, \ldots, 4$). 
The product $\prod_{m_i}$ runs over all the inverted bands at the high symmetry points $\Gamma_i$. 
$\zeta^{\textrm{IV}}_{m_i}(\Gamma_i)$ and $\zeta^{\textrm{IC}}_{m_i}(\Gamma_i)$ are the eigenvalues of the 
$180^\circ$-rotation for the inverted-valence (IV) 
and inverted-conduction (IC) bands respectively at the high symmetry point $\Gamma_i$. 
Same as in the main text, we use the eigenvalues of the inverted conduction bands in the topological material to
substitute the eigenvalues of the corresponding valence bands in the trivial insulator in the denominator.
As will be proved below, this ratio 
coincides with Eq.~\eqref{app:eq:weak_z2} and thus the parity of this mirror Chern number is identical to the weak topological index.

For a 2D system with D$_{2h}$ symmetry, the horizontal mirror reflection $\sigma$ is the product of the $180^\circ$-rotation along the 
normal direction ($C_2$) and the space inversion ($I$)
\begin{align}
\sigma=C_2\otimes I.
\end{align}
For bands with positive parities ($I=E$ where $E$ is the identity matrix), we find that $\sigma=C_2$ and 
thus $C_2$ and $\sigma$ have the same eigenvalues 
and eigenstates. This implies that at a high symmetry point $\Gamma_i$, for the Bloch wave $|\psi^+_m(\mathbf{k})\rangle$, which has 
mirror eigenvalue $+i$, it is also an eigenstate of $C_2$ with the same eigenvalue $\zeta=+i$. 
For bands with negative parity ($I=-E$), we have $\sigma=-C_2$ and thus $|\psi^+_m(\mathbf{k})\rangle$ has rotation eigenvalue 
$\zeta=-i$, opposite to the eigenvalue of the mirror reflection. In summary, the ratio between $\zeta$ for different bands at a high symmetry point is exactly 
the same as the ratio between parity eigenvalues ($\xi$) of the same bands. Therefore, we get
\begin{align}
(-1)^{C^{+}}=\prod_{i=1}^{4}\prod_{m_i} \frac{\zeta^{\textrm{IV}}_{m_i}(\Gamma_i)}{\zeta^{\textrm{IC}}_{m_i}(\Gamma_i)}
=\prod_{i=1}^{4}\prod_{m_i} \frac{\xi^{\textrm{IV}}_{m_i}(\Gamma_i)}{\xi^{\textrm{IC}}_{m_i}(\Gamma_i)}.
\label{app:eq:mirrorcandweak}
\end{align}
By comparing with Eq.~\eqref{app:eq:weak_z2}, we find that
\begin{align}
(-1)^{C^{+}}=(-1)^{\nu_{\textrm{weak}}}.
\end{align}
Therefore, we proved that for the D$_{2h}$ symmetry, (the parity of) the mirror Chern number obtained by the technique we use coincides 
with the weak Z$_2$ topological index, and thus it offers no additional insight beyond the Z$_2$ topological indices.

If we apply this technique to SmB$_6$, it is easy to notice that for the $k_x=k_y$ plane, which has D$_{2h}$ symmetry, the mirror Chern number
is 
\begin{align}
(-1)^{C^{+}}=\frac{\zeta_{\textrm{IV}}^{\textrm{SmB}_6}(X)}{\zeta_{\textrm{IC}}^{\textrm{SmB}_6}(X)}
=\frac{\xi_{\textrm{IV}}^{\textrm{SmB}_6}(X)}{\xi_{\textrm{IC}}^{\textrm{SmB}_6}(X)}=-1.
\end{align}
Here we used the fact that in SmB$_6$ at $X$ point, there is only one pair of inverted bands, which have opposite parities
$\xi_{\textrm{IV}}^{\textrm{SmB}_6}(X)=+1$  and $\xi_{\textrm{IC}}^{\textrm{SmB}_6}(X)=-1$.

\section{The $k_z=0$ plane}
For the $k_z=0$ plane (with D$_{4h}$ symmetry), as shown in the main text,
\begin{align}
(i)^{C^{+}_{\textrm{SmB}_6}}=\frac{\zeta_{\textrm{IV}}^{\textrm{SmB}_6}(X)}{\zeta_{\textrm{IC}}^{\textrm{SmB}_6}(X)}.
\end{align}
Utilizing the same arguments as we used in Eq.~\eqref{app:eq:mirrorcandweak}, 
we find
\begin{align}
(i)^{C^{+}_{\textrm{SmB}_6}}=\frac{\zeta_{\textrm{IV}}^{\textrm{SmB}_6}(X)}{\zeta_{\textrm{IC}}^{\textrm{SmB}_6}(X)}=
\frac{\xi_{\textrm{IV}}^{\textrm{SmB}_6}(X)}{\xi_{\textrm{IC}}^{\textrm{SmB}_6}(X)}=-1.
\end{align}
This conclusion implies that $C^{+}_{\textrm{SmB}_6}=2$ up to modulo $4$.

\section{The $k_z=\pi$ plane}
For the $k_z=\pi$ plane, the high symmetry points are $X$, $R$ and $M$. The first two have four-fold rotational symmetry 
along the normal direction ($z$) while $M$ has two-fold. Because of the D$_{4h}$ symmetry, the mirror Chern number 
of this plane is
\begin{align}
(i)^{C^{+}}=\prod_{m=1}^{N}(-1) \eta_{m}(X)\eta_m(R)\zeta_{m}(M).
\end{align}
By comparing with the trivial insulator with no band inversion, we find
\begin{align}
(i)^{C^{+}_{\textrm{SmB}_6,k_z=\pi}}=
\frac{\eta_{\textrm{IV}}^{\textrm{SmB}_6}(X)}{\eta_{\textrm{IC}}^{\textrm{SmB}_6}(X)}.
\end{align}
Here, $\eta_{\textrm{IV}}^{\textrm{SmB}_6}(X)$ and $\eta_{\textrm{IC}}^{\textrm{SmB}_6}(X)$ are the eigenvalues of 
the  $90^\circ$-rotation (along $z$) for the  inverted-valence (IV) and inverted-conduction (IC) bands at the momentum point 
$(0,0,\pi)$ respectively.  As will be proved below, this ratio is $+i$, and therefore the mirror Chern number is $+1$.

The eigenvalues of the $90^\circ$-rotation can be determined utilizing the representation of the space group, which can be easily 
constructed using the  character table and bases provided in group theory literature (e.g. Ref.~\onlinecite{Koster1963}). 
At $X$, $(0,0,\pi)$, possible representations for spin-$1/2$ particles are $\Gamma_7^{\pm}$ and $\Gamma_6^{\pm}$, 
which are the double-group representation of the little group at $X$ (D$_{4h}$). All these four representations
are two-dimensional representations and thus all the group elements can be written as $2\times 2$ matrices.
The matrices of the relevant operators are listed below. Here, we choose the basis such that the matrix of the mirror reflection is
diagonalized.
\begin{align}
U_{C_4}^{7\pm}=
\begin{pmatrix}
-\frac{\sqrt{2}}{2}+i\frac{\sqrt{2}}{2}&0\\
0&-\frac{\sqrt{2}}{2}-i\frac{\sqrt{2}}{2}
\end{pmatrix}
\end{align}
\begin{align}
U_{C_4}^{6\pm}=
\begin{pmatrix}
\frac{\sqrt{2}}{2}-i\frac{\sqrt{2}}{2}&0\\
0&\frac{\sqrt{2}}{2}+i\frac{\sqrt{2}}{2}
\end{pmatrix}
\end{align}
\begin{align}
U_{\sigma_h}^{7+}=
\begin{pmatrix}
-i&0\\
0&i
\end{pmatrix}
\end{align}
\begin{align}
U_{\sigma_h}^{6+}=
\begin{pmatrix}
-i&0\\
0&i
\end{pmatrix}
\end{align}
\\
\begin{align}
U_{\sigma_h}^{7-}=
\begin{pmatrix}
i&0\\
0&-i
\end{pmatrix}
\end{align}
\begin{align}
U_{\sigma_h}^{6-}=
\begin{pmatrix}
i&0\\
0&-i
\end{pmatrix}
\end{align}
The subscripts $C_4$ and $\sigma$ represent the $90^\circ$-rotation (along $z$) and the horizontal mirror reflection (about the $x-y$ plane) 
operators respectively. The superscripts $6+$, $6-$, $7+$, $7-$ mark the representations.

Using these matrices, it is easy to notice that if the two inverted bands have $\Gamma_7^{\pm}$ symmetry (as in SmB$_6$), 
the inverted valence band ($\Gamma_7^+$) has
\begin{align}
\eta_{\textrm{IV}}^{\textrm{SmB}_6}(X)=-\frac{\sqrt{2}}{2}-i\frac{\sqrt{2}}{2},
\end{align} 
and the inverted conduction band ($\Gamma_7^-$) has 
\begin{align}
\eta_{\textrm{IC}}^{\textrm{SmB}_6}(X)=-\frac{\sqrt{2}}{2}+i\frac{\sqrt{2}}{2}.
\end{align}
As a result, 
\begin{align}
(i)^{C^{+}_{\textrm{SmB}_6,k_z=\pi}}=\eta_{\textrm{IV}}^{\textrm{SmB}_6}(X)/\eta_{\textrm{IC}}^{\textrm{SmB}_6}(X)=+i.
\end{align}
If the two inverted bands are $\Gamma_6^{\pm}$ bands, the inverted valence band ($\Gamma_6^+$) has
\begin{align}
\eta_{\textrm{IV}}(X)=\frac{\sqrt{2}}{2}+i\frac{\sqrt{2}}{2},
\end{align} 
and the inverted conduction band ($\Gamma_6^-$) has 
\begin{align}
\eta_{\textrm{IC}}(X)=\frac{\sqrt{2}}{2}-i\frac{\sqrt{2}}{2}.
\end{align}
As a results, 
\begin{align}
(i)^{C^{+}_{\textrm{SmB}_6,k_z=\pi}}=\eta_{\textrm{IV}}^{\textrm{SmB}_6}(X)/\eta_{\textrm{IC}}^{\textrm{SmB}_6}(X)=+i.
\end{align}

\end{document}